\magnification=\magstephalf
\parindent=3pc
\parskip=3pt
\vsize=21.true cm
\hsize=16true cm
\baselineskip=17pt
\raggedbottom
\abovedisplayskip=3mm
\belowdisplayskip=3mm
\abovedisplayshortskip=0mm
\belowdisplayshortskip=2mm
\tolerance=500

\def\eg{{\it e.g.\ }}
\def\ie{{\it i.e.\ }}
\def \title#1    {\vbox to14mm{\vfil}\noindent{\uppercase{\srm#1}}
		 \ls\ls\ls}
\def \author#1   {\ls\parindent17.5mm{\uppercase{\srm#1}}}
\def \address#1  {\par\parindent17.5mm#1\parindent4mm}
\def \abstract   {\ls\ls\noindent{\srm ABSTRACT. \ }}

\def\ref{\noindent\hangindent.5in\hangafter=1}


\def\>{$>$}
\def\<{$<$}
\def\msun{M_\odot}
\def\ltsima{$\; \buildrel < \over \sim \;$}
\def\de{\partial} 
\def\simlt{\lower.5ex\hbox{\ltsima}}
\def\gtsima{$\; \buildrel > \over \sim \;$}
\def\simgt{\lower.5ex\hbox{\gtsima}}
%

\vskip 3truecm
\centerline{\bf OBSERVATIONAL CONSEQUENCES OF BARYONIC GASEOUS DARK MATTER} 
\vskip 1.5truecm
\centerline{\bf Yu. A. SHCHEKINOV}
\bigskip
\centerline{\it Department of Physics, Rostov State University,
Rostov on Don, Russia}
\medskip
\centerline{\it yus@rsuss1.rnd.runnet.ru}
\bigskip

\vskip 1.5truecm

\item{}{{\it Abstract.} Possible observational consequences of dark matter 
in the Galaxy in the form of dense molecular gas clouds -- clumpuscules 
of masses $M_c\sim 10^{-3}~\msun$ and radii $R_c\sim 3\times 10^{13}$ cm -- 
are considered. Recent models of the extreme scattering events -- refraction 
of radio-waves from quasars in dense plasma clumps in the Galactic halo -- 
definitely show on such clouds as possible dark matter candidate. We
argue that collisions of such clumpuscules are quite frequent: around 
$1-10~\msun$ a year can be ejected in the interstellar medium due to 
collisions. Optical continuum and 21 cm emissions from post-collisional gas 
are found to be observable. We show that clumpuscules can form around O 
stars HII regions of sizes $R\sim 30$ pc and emission measure 
$EM\simeq 20$ cm$^{-6}$ pc, and can also be observable in 
$H_\alpha$ emission. Evaporation of clumpuscules by external ionizing 
radiation can be a substantial mass source. From requirement that the total 
mass input on the Hubble time cannot exceed the luminous mass in the Galaxy, 
typical radius of clouds is constrained as $R_c<3.5\times 10^{12}$ cm, 
and their contribution to surface mass density as 50 $\msun$ pc$^{-2}$. 
It is argued that dissipation of kinetic energy of the gas ejected 
by such clouds can be an efficient energy source for the Galactic halo. 
}

\vfill\eject
\vskip 1.5truecm
\medskip
\centerline{\bf 1. Introduction}
\medskip
\par

Baryon dark matter has been supported observationally when microlensing 
events by massive compact halo objects (MACHOs) were detected [1--3]. 
The nature of MACHOs remains still unclear, although from general point of 
view stars of small masses, $M\sim 10^{-3}\msun$ (brown dwarfs, called 
often ``jupiters'') can provide gravitational lensing with observed 
characteristics, and on the other hand, being below the limit of 
hydrogen burning, they have too low luminosity to be seen. 
It is quite possible that microlensing events with observed frequency 
can be connected also with red dwarfs ($M\simgt 0.08~\msun$) with total 
mass exceeding the mass of luminous matter in the Galaxy [4]. 

Recently dense atomic and molecular clouds, confined by external
pressure or self-gravity, are 
discussed as an alternative possible reservoir of dark matter [5--14]. 
Contrary to brown dwarfs these clouds are assumingly cold 
(from 3 K [6] to 10 K [14]) and have 
negligible proper luminosity and are unseen even in infrared, their  
column densities are predicted to reach $\sim 3\times 10^{23}$ [11, 12] to 
$10^{25}$ cm$^{-2}$ [13], and they are optically thick in 21 cm. From
this point of view dark matter gas clouds are notoriously hard to
detect.

In [7] $\gamma$-radiation from interaction of cosmic rays with nuclei of
dark matter molecular clouds has been estimated and suggested to test the
presence of such clouds in the galactic halo. Very recently statistical
evidences for a large scale anisotropic excess in $\gamma$-rays were
presented [15], with the flux above 1 GeV close to the one predicted in
[7]. This can be considered in support of dark mass in the form of dense
cold gas. 

Additional evidences for the existence of dark matter gas clouds are 
connected with optical (non-gravitational) 
lensing of radio emission from quasars passing through atmospheres of 
ionized gas surrounding such clouds. Ten years ago dramatic short variations 
of fluxes (of several weeks to months duration) from compact radio quasars 
were discovered at wavelengths 3.7 and 11 cm -- the so-called 
extreme scattering events [16]. The origin of these variations was attributed 
to refraction by dense plasma clouds moving perpendicularly to the line of 
sight [16, 17]. From variations of radio fluxes lensing clouds are inferred 
to have characteristic sizes of a few AU and electron 
densities of $n_e\simeq 10^3$ cm$^{-3}$. The internal gas pressure 
$n_e T_e\simgt 10^7$ K cm$^{-3}$ exceeds pressure in the surrounding gas 
by 3-4 orders of magnitude, and therefore refracting plasma clouds must be 
transient structures of several months life time, 
which is far from being enough to explain observed frequency of the 
extreme scattering events [16, 17]. Only recently this 
difficulty has found a solution [18] with the assumption that lensing plasma 
can be connected with extended HII atmospheres around cool dense 
self-gravitating molecular clouds as was suggested in [9]. 
In the framework of this model high interior pressure provides 
steady outflow of ionized gas from the HII cloud atmosphere, and 
thus the density distribution in lensing plasma regions is kept 
time-independent. The individual cloud masses are high enough to support 
photoionization mass outflow on galactic time scales. 
In [18] these clouds were identified with the dense molecular 
clumpuscules described in [5, 6]. This allowed the authors to
reproduce with good accuracy the dual-wavelength 
(3.7 and 11 cm) light-curve of the quasar 0954+658, while all the 
previous models had difficulties with reproducing this.  
Quite recently, the idea that such 
dense molecular clouds contribute significantly to dark matter has 
obtained further support: it was shown that they act as gaseous
(non-gravitational) lenses for stellar light, which in turn can result
in the observed excess of microlensing events, compared to the expected
from stars and stellar remnants toward the LMC [19]. 
In this paper we consider possible dynamical and observational 
consequences of the existence of dark matter in the form of 
dense molecular clouds. (Through the paper we will refer these clouds 
to as DM-clouds, or following [13], clumpuscules). 

In \S 2 we will discuss general properties of the population of DM-clouds in 
the Galaxy which can be inferred from the results of [18]; in \S 3 we 
describe qualitatively dynamics of collisions of DM-clouds; 
in \S 4 we estimate optical emission from colliding clumpuscules; 
in \S 5 we estimate 21 cm emission from clumpuscules and their debris 
after collisions; in \S 6 we consider $H_\alpha$ emission from 
ionized atmospheres of DM-clouds; in \S 7 possible implications for 
the Galactic halo gas is discussed; \S 8 contains discussions of the 
results and conclusions. 

\bigskip
\centerline{\bf 2. General properties of molecular DM-clouds.}
\medskip
\par
In [5, 6] molecular DM-clouds were assumed 
to have high enough total mass (about one order of magnitude in excess of 
the mass of visible matter), quite low proper luminosity, and life time 
comparable with the Hubble time. It follows 
then that individual masses of these objects must be close to the Jovian
mass, $M_c\sim 10^{-3}~\msun$, and densities, temperatures and sizes such 
to make them marginally gravitationally stable:  
$n_c\sim 10^{11}$ cm$^{-3}$, $T_c\sim 3$ K, $R_c\sim 10^{14}$ cm. 
The interpretation of the extreme scattering events suggested in [18] 
leads to very close value for the radius of a DM-cloud: 
the duration of these events around 2 months suggests that 
$R_c\simlt 3\times 10^{14}$ cm for the transverse velocity of DM-clouds 
restricted from above by the escape velocity for the Galaxy 
500 km s$^{-1}$. The covering factor of the clouds is estimated from 
the flux monitoring data [16] as $f\sim 5\times 10^{-3}$. It allows, in turn, 
to estimate the contribution of DM-clouds to the total surface density 
of the Galaxy as [18]: $\Sigma \sim fM_c/\pi R_c^2$, where $M_c$ 
is found from hydrostatic equilibrium $M_c\sim 
kT_c R_c/Gm_p$. It gives for $T_c\simgt 3$ K the surface density 
$\Sigma \simgt 10^2~\msun$ pc$^{-2}$ [18], which is close to the value 
estimated from dynamical arguments [20]. 

The characteristic free-path length of DM-clouds with respect to cloud-cloud 
collisions is of $r_0/f$, where $r_0$ is the characteristic size of 
the region occupied by clouds. For $r_0\sim 10$ kpc, one gets for 
characteristic time between collisions $\simgt 10^{10}$ yr. Cloud destruction 
by the external UV radiation is slow enough with similar characteristic time, 
due to their high masses and densities. Moreover, contact interactions 
of DM-clouds with interstellar clouds are inefficient because of 
high differences in densities, and thus one can conclude (see [18]) that 
the population of DM-clouds does not change significantly over the Hubble 
time. At the same time, even so inefficient destruction of DM-clouds can 
result in dynamically important consequences because their total mass in 
the Galaxy is large. 

We will assume for specifity the distribution of DM-clouds in the form  

$${\cal N}={{\cal N}_0\over 1+(r/r_0)^2},\eqno(1)$$
which fits the singular isothermal sphere at $r>r_0$ (through the paper
$r_0$ is regarded as a free parameter); 
here ${\cal N}_0$ is the number of clouds per unit volume in the   
galactic center, $r$ is the galactocentric radius. 
The collision rate per unit volume is then 

$$\nu=\pi R_c^2{\cal N}^2v_c,\eqno(2)$$ 
where all clouds are assumed to have equal characteristic velocity $v_c$ 
independent of their position in the Galaxy. In what follows we accept in 
numerical estimates $R_c=3\times 10^{13}$ cm and $v_c=250$ km s$^{-1}$  -- 
two times less than the escape velocity -- unless other values specified. 
The total rate of collisions in the Galaxy is 

$$\nu_G=4\pi \int\limits_0^\infty \nu(r) r^2 dr=\pi^3 R_c^2 r_0^3 
{\cal N}_0^2 v_c,\eqno(3)$$ 
which is connected with the characteristic covering factor of DM-clouds 
$f\sim \pi{\cal N}_0R_c^2 r_0$ by equation 

$$\nu_G\sim {4\over \pi}{r_0v_c\over R_c^2}f^2.\eqno(4)$$
For adopted values of $R_c$, $v_c$ and $f$ this gives 
$\nu_G\sim 10^{-26}r_0$ s$^{-1}$, or 
$\nu_G\sim (10^3-10^4)$ yr$^{-1}$ for $r_0=1$ kpc and $r_0=10$ kpc, 
respectively; the total number of clumpuscules in the Galaxy can reach 
$N\sim 10^{13}-10^{15}$. (Note, that for given characteristic length of clouds 
distribution $r_0$ and fixed total number of clouds in the Galaxy $N$, 
the surface density from clumpuscules is proportional to their 
radius $\Sigma\sim kT_cNR_c/\pi G m_p r_0^2$, however, for fixed 
current covering factor it is inversely proportional to $R_c$: 
$\Sigma\propto fR_c^{-1}$.) Therefore, the amount of material of molecular 
clumpuscules processed by collisional shocks is 
$\sim \nu_G M_c\sim 1- 10 ~\msun$ yr$^{-1}$. This value is comparable or 
exceeds typical rates of matter processing in such galactic events as 
star formation, heating of the interstellar gas by supernovae shocks, 
accretion of galactic halo gas to the plane. One can expect thus that 
collisions of DM-clouds have important consequences for the interstellar 
gas and might be observed. 

\bigskip
\centerline{\bf 3. Collisional dynamics of molecular DM-clouds.}
\medskip
\par 
In this Section we describe qualitatively dynamics of collisions of DM-clouds. 
Detailed description of numerical results will be given elsewhere. Let us
consider a head-on collision of two equal clouds, and neglect self-gravity 
effects. In the center-of-mass frame a contact discontinuity 
forms in the symmetry plane, from which two shocks propagate outward. 
Characteristic duration of the collision, \ie time needed for the shock to 
reach outside boundary of the cloud, is 
$\tau_s\sim R_c/v_c \sim 0.4$ yr for adopted parameters. 
Due to high density behind the shock, $n_s=4 n_c\sim 
4\times 10^{12}$ cm$^{-3}$, radiation cooling time in bremsstrahlung 
processes is only $\tau_R\sim 3$ s. However, hard radiation emitted 
by shocked gas is absorbed on scales 
$\sim 10^6$ cm, and after time interval $\sim 10^5$ s, 
when the shock wave covers around 1 - 2 \% of the cloud mass, it becomes 
ionized $x\sim 1$ and Thomson optical depth increases to $\tau_T\sim 
20$. This circumstance allows us to neglect radiative cooling. 

Due to high postshock pressure, $P_s\sim \rho_c v_c^2
\sim 10^3$ erg cm$^{-3}$ ($\sim 10^{19}$ K cm$^{-3}$), 
a strong axial gas outflow along the symmetry plane forms immediately 
after collision, which decreases pressure and temperature behind the 
shock, and moreover it scatters a substantial mass fraction of 
clouds during the collision $t\sim \tau_s$. 
In [21] this effect was described first and then confirmed 
in a sequence of papers [22-26]. The fraction   
of scattered mass for diffuse interstellar clouds dominated by radiative 
cooling was estimated in [21] as 25 \%. In radiationless case this 
fraction can be somewhat larger due to higher pressure. (One should stress, 
however, an approximate character of these estimates -- recent simulations 
with high precision [26] have demonstrated that at late stages after 
collision, $t>10-15$ dynamical times, Rayleigh-Taylor 
instability develops to form multiple clumps and filaments, and 
violates strongly the integrity of post-collisional clouds. Thus, 
the fraction of escaped gas can be higher.) 

After the shock wave reaches external boundary of the cloud, a rarefaction 
wave starts to propagate inward resulting in expansion of the cloud with 
velocity equal to the local sound speed (see [21]). At  
$t\simlt \tau_s$ the rarefaction wave reaches the symmetry plane and 
reflects. When the reflected 
wave reaches external boundary of the cloud, gas pressure in outer regions 
relaxes already to external pressure, and thus the secondary reflected 
wave is a compression wave. These wave motions 
result in separation of clouds with final pressure relaxed to the external 
value. Since viscosity behind shock waves works to increase entropy, cloud 
collisions are dissipative even in radiationless case. This means that 
after separation of clouds pressure equilibrium establishes at gas density 
less than the initial one: the new equilibrium corresponds to the Poisson 
adiabat with entropy larger than the initial value. 

Self-gravity effects on collisional dynamics are weak, and  
are only important for unperturbed clouds (\ie before 
collision), as they are assumed to be marginally stable [5, 6]. 
After collision thermal energy of clouds becomes larger than their 
gravitational energy due to dissipative increase of entropy at the shock, 
and separated clouds expand freely up to pressure equilibrium with
ambient medium. At late stages clouds can get 
transparent to external heating sources, and their subsequent expansion 
settles in an isothermal regime, and finally they mix with surrounding gas. 

An important qualitative difference of shocked gas in colliding clumpuscules 
from that described in [21 -- 26] is connected with the fact that due to 
high relative velocity of clumpuscules the postshock gas is dominated by 
radiation pressure. As mentioned above, Thomson optical depth of the 
shocked gas reaches unity quickly, and thus gas and radiation behind the 
shock set in equilibrium. The postshock pressure is then 

$$P={{\cal R}\over \mu}\rho T+{4\over 3}{\sigma_{_{SB}} T^4},\eqno(4)$$
where ${\cal R}$ is the gas constant, $\sigma_{_{SB}}$ 
[erg cm$^{-3}$ 
K$^{-4}$], the Stefan-Boltzmann constant. On the other hand, 
$P=\rho_0 D^2(1-\eta)$, where $\eta^{-1}=\rho/\rho_0$ is the compression 
factor, $D$, the shock velocity in the center-of-mass frame. 
It can be readily shown that the solution of eq. (4) at 
$v_c=250$ km s$^{-1}$ is $T\simeq 2.8\times 10^4$ K, and the contribution of 
gas pressure [first term in eq. (4)] is $\sim 0.01$ for $n_c=10^{12}$
cm$^{-3}$. Fractional ionization (determined by Saha equation) is $x=1$ 
(more precisely $1-x\simeq 6\times 10^{-7}$). Due to radiation diffusion the 
shock front is smoothed (see [27]) over scales of several free-path lengths 
of photons, in our case $\sim 10^{-2}R_c$. From this point of view, 
qualitative dynamics of shocks and subsequent separations of clouds after 
collision is similar to that described in [21]. The sound speed behind the 
shock is 

$$c_s={2\over 3}{\sigma_{_{SB}}^{1/2}T^2\over \rho^{1/2}},\eqno(5)$$ 
and for adopted parameters $c_s\simeq 0.7\times 10^7$ cm s$^{-1}$. 

\bigskip
\centerline{\bf 4. Optical emission.}
\medskip
\par
The postshock temperature is 

$$T_s\simeq 2.8\times 10^4 \left({v_c\over 250~{\rm km~s^{-1}}}\right)^{1/2} 
~~{\rm K},$$ 
for $v_c\simgt 15$ km s$^{-1}$ (for $n_c\sim 10^{12}$ cm$^{-3}$), at lower 
velocities the contribution of radiation to pressure behind shock is 
smaller than gas pressure, and temperature is determined by 

$$T_s\simeq 2.5\times 10^6\left({v_c\over 250~{\rm km~s^{-1}}}\right)^2~~
{\rm K}.$$ 
Thomson optical depth after ionization of the preshock gas reaches 
$\tau_T\sim \sigma_T R_c n_c\sim 20$. The luminosity of the colliding 
DM-clouds is determined by diffusion of photons from the hot interior and  
is equal to 

$$L_h\simeq \sigma_{_{SB}}\tau_T^{-1} cT^4 S,$$ 
where $S\sim 2\pi R_c^2$ is the emitting area. For adopted parameters  
it gives $L_h\sim 10^{38}$ erg s$^{-1}$. The duration of hot 
stages is determined by gas expansion after shock waves reach external 
boundaries of the clouds and the rarefaction waves start to propagate 
inward: $t_d\sim R_c/3 c_s\sim 2\times 10^6$ s. Thus, the total energy loss 
by radiation is $L_h t_d\sim 2\times 10^{44}$ erg, \ie about 30 \% of the 
total kinetic energy of clouds. 

Observationally colliding DM-clouds at these stages are short-lived transients 
(with the life time $t\sim t_d\sim 10^d$), with temperature and luminosity 
typical for massive stars. The total number of such 
objects in the Galaxy is $\nu_G t_d\sim 80(r_0/1~{\rm kpc})$. 
Only small their fraction $\sim 2 \pi^{-1} {\rm arctg}(r/ r_0)$, is located 
inside a region with galactocentric radius less than $r$, and can be obscured 
by dust in the galactic central parts: this fraction inside $r<1$ kpc is 
0.5 for $r_0=1$ kpc (\ie the number of unobscured objects is $\sim 40$), 
and only 0.06 for $r_0=10$ kpc. Their characteristics 
(for $v_c\geq 100$ km s$^{-1}$ and $n_c\sim 10^{12}$ cm$^{-3}$) scales as 
the total number of clouds at hot stage 
$$N_h={6\over \pi}\left({1-\eta\over \eta}\right)^{1/2}f^2{r_0\over R_c}
\simeq 80~\left({r_0\over 1~{\rm kpc}}\right)\left({R_c\over 3\cdot 10^{13}~ 
{\rm cm}}\right)^{-1},$$ 
cloud luminosity 
$$L_h={2\pi \eta^2\over 1-\eta}{m_pv_c^2R_c c\over \sigma_T}\simeq 
7.5\times 10^{37}\left({R_c\over 3\cdot 10^{13}~{\rm cm}}\right)
\left({v_c\over 250~{\rm km~s^{-1}}}\right)^2~~{\rm erg~s^{-1}},$$
temperature 
$$T_h\simeq \left[{\eta^2\over 1-\eta}{\rho_0 v_c^2\over \sigma_{_{SB}}}
\right]^{1/4}\simeq 2.8\times 10^4\left({n_c\over 10^{12}~{\rm cm}^{-3}}
\right)^{1/4}\left({v_c\over 250~{\rm km~s^{-1}}}\right)^{1/2}~{\rm K},$$ 
here $\eta \simeq 1/6$. 
Possible detection of such objects (or contrary, non-detection at given 
threshold) might allow to infer (or restrict) characteristics of DM-clouds 
and estimate their contribution to the surface density of the Galaxy.

\bigskip
\centerline{\bf 5. Emission in 21 cm.}
\medskip
\par
Estimates of emission in 21 cm after collision of clouds can be 
obtained assuming each cloud to expand isotropically after the primary shock 
reaches the external surface of the cloud. At such an assumption, initial 
stages can be described as an expansion with $\gamma=4/3$ 

$$P=P_1\left({\rho\over \rho_1}\right)^{4/3},$$ 
where subscript 1 refers to variables at the moment of maximal compression: 
$P_1=4\sigma_{_{SB}}T_1^4/3$, $T_1=2.8\times 10^4$ K, $\rho_1=\rho_c/\eta$. 
At stages when Thomson optical depth is large, $\tau_T >1$, both radiation 
and gas temperatures vary as $T\propto R^{-1}$. For a quasi-spherical 
expansion with gas density varying as $\rho\propto R^{-3}$, both gas 
($P_g$) and radiation ($P_\gamma$) pressure vary as $\propto R^{-4}$, thus 
the initial ratio $P_g/P_\gamma\simeq 0.05$ is kept up to stages 
when electron density decreases substantially and $\tau_T$ becomes less 
than one. It can be readily shown that $\tau_T=1$ is reached when temperature 
decreases to $T_2=5.8\times 10^3$ K. At this moment 
$n_2=0.016~n_1\simeq 10^{11}$ cm$^{-3}$, $x=0.12$, 
and column density of neutral hydrogen is 
$N_2(HI)\simeq 6\times 10^{23}$ cm$^{-2}$. Note, that hydrodynamical 
instabilities in post-collisional gas, found in recent numerical 
simulations, produce non-homogeneous density distribution in clouds and 
their fragments, however, the density contrast between different regions 
does not exceed factor of 3-4 at times $t>50-60$ dynamical times [26], 
which in our case corresponds to $t\sim 10^6-10^7$ s. This means, that 
our description based on the assumption of smooth structure of 
post-collisional clouds, gives correct estimates by the order of magnitude. 

Subsequent expansion is adiabatic with $\gamma=5/3$, and at 
$n_3=10^6$ cm$^{-3}$ gas temperature is $T_3\simeq 3$ K. The duration of this 
stage is $\sim R_3/c_s\sim 2\times 10^{11}$ s, where $R_3\sim 
(n_1/n_3)^{1/3}R_c$ is the characteristic size of the cloud 
corresponding to density $n=n_3$: for adopted numbers $R_3\sim 10^{15}$
cm. Note, that possible existence of such cold cloudlets in the gaseous
galactic halo is argued in [28] from analysis of turbulent motions in
the neutral halo gas. Further the cloud expands isothermally with temperature 
$T\simeq 3$ K, which is supported by deactivating inelastic collisions of 
hydrogen atoms with electrons $H(F=1)+e~\to H(F=0)+e'$, with a rate 
$q_{10}=3\times 10^{-11}~T^{1/3}$ cm$^3$ s$^{-1}$ [29] -- the  
corresponding characteristic time $t_{10}\sim 2\times 10^4$ s. 
Isothermal phase stops when gas density decreases to $n_4\sim 3\times 10^3$ 
cm$^{-3}$ and the cloud comes to pressure equilibrium with the ambient 
interstellar gas. At this state heating by the background X-ray emission 
gets important with characteristic time 
$t_X\sim 4\times 10^{11}$ s (the heating rate $\Gamma_X\simeq 10^{-27}$ 
erg s$^{-1}$ H$^{-1}$ is taken from [30]). Subsequently, the cloud expands 
slowly with charateristic time $t_X$, and becomes transparent to 21 cm 
when HI column density falls to $N(HI)\sim 10^{19}$ cm$^{-2}$ -- 
it occurs at $t=t_{HI}\sim 10^{13}$ s when $n=n_5\sim 10^3$ cm$^{-3}$. 
Apparently, at $t>t_{HI}$ the cloud mixes with the interstellar gas. The 
total number of such clouds in the Galaxy is estimated as 
$\nu_Gt_{HI}\sim 3\times(10^8-10^{9})$, and their total mass as 
$3\times(10^5-10^6)~\msun$ for $r_0=1-10$ kpc, respectively. 
A distinctive feature of these clouds is their high velocity dispersion which 
is close to the dispersion of DM-clouds $v_c$ (for DM-clouds with Maxwellian 
velocity distribution the mean center-of-mass velocity for pairs of clouds is 
$\sim \sqrt{3/2}v_c$) -- much larger than the local velocity dispersion of 
the interstellar HI. This circumstance can be used for identification 
of such clouds. It should be mentioned, however, that when cloud fragments 
and debris mix with surrounding gas, a fraction of their kinetic energy 
converges to thermal energy due to viscosity, and thus a fraction of their
mass will be heated up to high temperatures and thus will be unseen 
in 21 cm. To evaluate this effect, note that drag force for a cloud moving 
through interstellar gas is proportional to ram pressure 
$\propto \rho_i v_c^2 R_c^2$, even for subsonic clouds [31]. 
In these conditions the length scale for velocity decay is 
$\ell_d \sim \delta R_c$, where $\delta$ is the density contrast between 
cloud and intercloud gas [32]. At stages when fragments become transparent 
to 21 cm gas density is of order $n_5\sim 10^3$ cm$^{-3}$ and radius 
$R_4\sim 10^{16}$ cm. 
Thus, cloud velocity decays on scales $\ell_d \sim 30-300$ pc for density 
of ambient gas $n_i\sim 0.1-0.01$ cm$^{-3}$, respectively. The corresponding 
characteristic time $t_d\sim 3\times 10^{12-13}$ s is comparable to 
$t_{HI}$. Therefore, although dynamics of cloud fragments at late stages can 
depend on environment crucially, time interval for fragments to be seen in 
21 cm is close to $t_{HI}$. At these stages, before mixing with the ambient
gas, fragmets form a network with covering factor of $\sim 0.01-0.3$ and
resemble the tiny-scale atomic structures described in [33]. 

DM-clouds of smaller radii, $R_c\simlt 3\times 10^{12}$ cm, 
have small Thomson optical depth and due to putchy density distribution
generated after collision by hydrodynamical instabilities, [26], they
are apparently transparent. In this case, as pointed in [34], 
after collison they quickly lose their energy radiatively and escape the
hot phase. However, subsequent expansion, starting from stages when
temperature has fallen to several thousands, is qualitatively 
similar to the described above, though it is less rapid since at given 
pressure clouds have larger densities. 


\bigskip
\centerline{\bf 6. $H_\alpha$ emission from isolated DM-clouds.}
\medskip
\par
By isolated clouds we mean clouds escaping collisions. 
Gas temperature in extended ionized gas around DM-clouds is of order 
$T\sim 10^4$ K, which for electron density $n_e\sim 10^3$ cm$^{-3}$ produces 
pressure by 3-4 orders of magnitude in excess of typical interstellar pressure. 
In such conditions ionized gas expands freely with the 
velocity $u_0=\sqrt{2/\gamma-1}c_0$, where $c_0\simeq 10$ km s$^{-1}$ is the 
sound speed. Therefore, ionized atmospheres of DM-clouds 
are in steady outflow regime with density and velocity distributions 
described by the equations (see, \eg [35]) 

$$\rho u r^2=\rho_0 u_0 R_c^2,$$ 

$${u^2\over 2}+{P_0\over \rho_0} \ln {\rho\over \rho_0}={u_0^2\over 2},$$ 

$$P=P_0{\rho\over \rho_0},$$ 
which give for density distribution the solution 

$$\left({\rho\over \rho_0}\right)^2 \left[
1-{2\over 5}\ln {\rho\over \rho_0}\right]=\left({R_c\over r}\right)^4.
\eqno(6)$$ 
The equation of ionization equilibrium 

$$4\pi \alpha_r R_c^3n_{e0}^2\int\limits_1^\infty\left({\rho\over \rho_0}
\right)^2\zeta^2 d\zeta=4\pi R_c^2 J,$$  
with $n_e\propto \rho$ taken from equation (6) determines the boundary value 
$n_{e0}$ at $r=R_c$ as 

$$n_{e0}\simeq \sqrt{{2 J\over \alpha_r R_c}},$$ 
where $\zeta=r/R_c$, and $J$ [cm$^{-2}$ s$^{-1}$] is the intensity  
of ionizing photons at Lyman edge, $\alpha_r$, the recombination rate. 
The corresponding emission measure is 

$$EM=\int\limits_{R_c}^\infty n_e^2 dr=0.236~n_{e0}^2R_c=
1.8~{\rm cm^{-6}~pc},$$ 
here we adopted $J$ obtained in [36] at high galactic latitude
$J=J_0\simeq 10^6$ cm$^{-2}$ s$^{-1}$ sr$^{-1}$, or the intensity 
of $H_\alpha$ photons of $\sim 1$ R 
($=10^6/4\pi$ phot cm$^{-2}$ s$^{-1}$ sr$^{-1}$). 
This number is by an order of magnitude higher than obtained in [37] for 
high-velocity HI clouds. However, because of small angular sizes $H_\alpha$ 
fluxes from distant DM-clouds at $d>1 $pc with angles $\Delta \phi< 0'.2$ 
are too weak to be detected at present. At the same time, clouds at 
distances $d\sim 0.03$ pc with angular sizes $\Delta \phi \sim 7'$ could be 
seen in $H_\alpha$. For adopted parameters and $r_0=1$ kpc, a 
volume $d^3\sim (0.03)^3$ pc$^3$ in solar vicinity contains in average one 
DM-cloud; for $r_0=10$ one cloud is contained in a volume with radius 
$0.07$ pc -- at such distance cloud angular size is only $\sim 3'$, and 
$H_\alpha$ flux decreases by factor of 5. 

Much larger emission measures have clumpuscules located in close vicinity of 
hot stars where the intensity of ionizing photons is higher than the 
background value $J\gg J_0$. For O7 stars this condition holds at 
$r< 30$ pc. The emission measure increases proportionally to 
$J/J_0$, and thus a DM-cloud at distance $\sim 1$ pc from hot star can be 
seen as an ultracompact $H_\alpha$ nebula as pointed in [12, 18]. 
However, because of small angular sizes 
($\sim 0''.1$ for a cloud at $\sim 100$ pc from Sun), expected fluxes are 
too low: $F_{H\alpha}\sim 3\times 10^{-3}$ phot cm$^{-2}$ s$^{-1}$ from 
a single cloud within 1 pc near an O7 star. In general, the total 
number of $H_\alpha$ photons emitted by a single cloud is 

$$L(r)\simeq 0.025 {{\cal L} R_c^2\over r^2}~{\rm phot~ s^{-1}},\eqno(7)$$
where ${\cal L}$ [phot s$^{-1}$] is the number of ionizing photons from 
a star, $r$, the distance to the star. A distance from the star where 
the background UV flux is equal to the flux produced by star is 

$$R_m=\sqrt{{{\cal L}\over 4\pi J_0}},$$  
$R_m\simeq 30$ pc for an O7 star. Total number of $H_\alpha$ photons emitted 
by all DM-clouds inside $r\leq R_m$ is then determined as 
$L_m=4\pi {\cal N}\int_0^{R_m}L(r)r^2 dr$ and is equal 

$$L_m=0.3 {{\cal L}^{3/2}R_c^2 {\cal N}\over \sqrt{4\pi J_0}}~
{\rm phot~s^{-1}}.$$ 
For adopted numerical values this gives $\simeq 7\times 10^{44}$ 
phot s$^{-1}$, however the corresponding surface brightness is extremely 
small -- the equivalent emission measure is only 
$EM\simeq 0.0025$ cm$^{-6}$ pc. 

One of possible observational manifestations of DM-clouds is connected 
with the fact that outflowing ionized gas fills the space around an O
star, and if the outflow rate is high enough it can support high density of 
electrons and can be observed as a luminous halo. In order to estimate this 
effect let us write the approximate equation of mass balance at distance $r$ 
from an ionizing star as 

$${\de \over \de r}(\rho u)={\cal N}\dot M_c(r) r^2,\eqno(8)$$
where $\dot M_c=2\pi R_c^2 m_pn_{e0}(r)u_0$ is the mass loss rate by a 
single cloud (factor $2\pi$ accounts that only one side of the cloud is 
ionized by the star), $n_{e0}(r)=4.8\times J^{1/2}R_c^{-1/2}$, 
(thus $n_{e0}\propto r^{-1}$ in virtue of $J\propto r^{-2}$). Note, that 
a volume with $r\sim 10$ pc contains $\sim 10^6$ to $\sim 10^7$ clouds,  
and this substantiate the assumption of steady state mass
and energy balance. In this case, integration over $r$ gives 

$$m_p n u={\cal N}\dot M_c(r) r,$$ 
or finally 

$$n=8.5\times 10^6~ {\cal L}^{1/2}R_c^{3/2}{\cal N}{u_0\over u}.\eqno(9)$$ 

To complete description, Bernoulli and energy equations must be written. 
As we will show below, gas pressure in regions filled by gas from 
HII-atmospheres is much higher than pressure in surrounding 
interstellar gas, and thus one can assume the outflow velocity $u$ to be 
independent of $r$ and equal with the factor $\sqrt{2/\gamma-1}$ to the 
sound speed, which in turn is determined by energy equation. It is readily 
seen that in this case $n=$const. Energy budget of the gas is determined by 
heating radiation from central star and by dissipation of high-velocity 
turbulent motion of the gas lost by DM-clouds from one side, and radiation 
cooling from the other. The mean velocity of the gas lost by clouds relative 
to intercloud gas is approximately equal to clouds velocity $v_c$, and thus 
the heat released due to dissipation of kinetic energy of evaporated gas 
can be considerable, $\sim 10^{14}$ erg g$^{-1}$ (the corresponding 
temperature $T\sim 10^7$ K). At such assumptions energy equation can be 
written as 

$$H_{UV}(T)n^2+{1\over 2}{\cal N}\dot M_cv_c^2=\Lambda(T) n^2,\eqno(10)$$ 
where $H_{UV}(T)n^2$ is the heating rate by UV radiation from central star, 
$H_{UV}(T)=\alpha_r(T)\epsilon_0$, $\epsilon_0$, the mean energy of 
photoelectrons (see [29, 38]), $\Lambda(T)$, the radiation loss rate. 
Assuming DM-clouds to be of primordial chemical composition [6, 13], one 
can accept that radiation cooling in interval $T=10^4-10^6$ K is determined 
by recombinations and equal to 
$\Lambda(T)\simeq 9.5\times 10^{-17}T^{-3/2}$ erg cm$^3$ s$^{-1}$ at  
$T=10^4-6\times 10^4$ K and $\simeq 3\times 10^{-21} T^{-1/2}$ at 
$T=6\times 10^4-10^6$ K (see [29]). Eliminating $n$ from (9) and (10) and 
assuming $u=\sqrt{2/\gamma-1}c_s$, we find that temperature varies from 
$T\simeq 10^4$ K at $r=1$ pc to $T=3\times 10^4$ K at $r=30$ pc. Therefore, 
one can accept in estimates $T=10^4$ K in the whole range of $r$, 
which gives $u\simeq 20$ km s$^{-1}$. Substituting $u$ in (9)  we arrive at 

$$n\simeq 4.2\times 10^6 {\cal L}^{1/2}R_c^{3/2}{\cal N},\eqno(11)$$ 
and

$$EM\simeq 1.2\times 10^{-6} 
{\cal L}R_c^2{\cal N}^{4/3}~{\rm cm^{-6}~pc},\eqno(12)$$
which gives $n \simeq 0.65$ cm$^{-3}$ and $EM\simeq 20$ for adopted 
parameters and $r_0=10$ kpc. The surface brightness in $H_\alpha$ is 
large enough $I_{H\alpha} \simgt 8$ R, and thus such HII halos can be observed. 
%

\bigskip
\centerline{\bf 7. Contribution of DM-clouds to energy balance 
of the Galactic corona.}
\medskip
\par

DM-clouds spend most time in the halo, where their ionized atmospheres 
are supported by the background UV radiation [18]. The mass loss rate 
by a single cloud is at such conditions 

$$\dot M_c=4\pi m_p\sqrt{{2J_0\over \alpha_r}}u_0R_c^{3/2},\eqno(13)$$
or numerically $\dot M_c\sim 10^{-13}~\msun$ yr$^{-1}$. 
The evaporating gas loses its relative velocity due to viscous forces on 
characteristic length of $\ell_d \sim \delta R_c$ (see \S 5), 
where radius of ionized atmosphere is accepted to be of order of 
cloud radius. For mean densities of halo gas $n_h\sim 10^{-3}$ cm$^{-3}$ and  
of ionized atmosphere $n_e\sim 10^3$ cm$^{-3}$ this gives 
$\ell_d\sim 10^6~R_c\sim 10$ pc which is much less than typical galactic 
scales. Thus, the energy input rate from kinetic energy of evaporating  
atmospheres is order of $H_c\sim \dot M_c{\cal N}v_c^2/2\sim 
2.5\times 10^{-25}$ erg cm$^{-3}$ s$^{-1}$, which gives 
$\int H_c dz\sim 2.5\times 10^{-3}$ erg cm$^{-2}$ s$^{-1}$ per unit area 
(here ${\cal N}\sim f/\pi R_c^2r_0$, see \S 2, $R_c=3\times 10^{13}$ cm and 
$r_0=10$ kpc are substituted). For comparison, the total energy input rate 
from SNe II is $\sim 2\times 10^{-4}$ erg cm$^{-2}$ s$^{-1}$ for SNe II 
galactic rate $1/30$ yr$^{-1}$. The radiation loss rate for gas with 
temperature $T\sim 10^6$ K and density $n\sim 10^{-3}$ cm$^{-3}$ is order 
of $\Lambda(T)n^2\sim 3\times 10^{-29}$ erg cm$^{-3}$ s$^{-1}$. In 
principle, observations show that the halo gas has extremely non-homogeneous 
density and temperature distributions. In this case, one can expect the 
total rate of radiation energy losses to be higher than the above estimate. 
In particular, in regions with $n\sim 10^{-2}$ cm$^{-3}$ and 
$T\sim 10^5$ K radiation takes away $\sim 10^{-25}$ erg 
cm$^{-3}$ s$^{-1}$. However, energy input rate from evaporating clumpuscules 
$H_c$ is much larger than possible energy losses. 

In this connection, one should mention a problem of energy balance in 
galactic halo gas, still far from being understood well. It is commonly 
believed that clustered SNe II in OB associations are the source which 
replenishes energy losses in the halo. Detailed calculations show however, 
that special conditions are needed to form vertical tunnels able to 
conduct energy from the disk to halo. It is shown in [39] that hot gas from  
a clustered SNe explosion can reach heights $z\sim 1-3$ kpc only if 
$\sim 100$ SNe exploded at distance $z=80$ pc from the plane. Definitely, 
such events are rare in the Galaxy. Quite recently additional results 
have been obtained which exacerbates the problem. Based on recent all sky 
survey of HI gas, soft X-ray radiation, high energy $\gamma$-ray emission and 
the 408 MHz survey, it has been found in [28] that radial distribution of 
halo gas has scale length $R_h\sim 15$ kpc. At the same time, galactic 
supernovae have different radial distributions: either exponential with 
radial scale length $\sim 4$ kpc for SNe I, or a ring with radius 
$\sim 5$ kpc and width $\sim 4$ kpc for SNe II. One might assume that SNe 
explosions in galactic central regions eject hot gas, which then fills 
the halo  in radial direction over galactic scales $R\sim 10-15$ kpc. 
However, characteristic time for hot ejected gas to expand in radial 
direction out of solar circle is of order $\sim 10^8$ yr, which is 
comparable to its radiative cooling time $3-10\times 10^7$ yr. At the 
same time, the problem can have a simple solution if transfer of kinetic 
energy of evaporating material of DM-clouds to thermal energy is taken 
into account. Note, that this mechanism can be comparable (and out of 
solar circle can exceed) to SNe explosions, even if the total number of 
DM-clouds is smaller than the value accepted in previous 
sections -- the rate $H_c$ decreased by factor of 10 is still of the same
order as the SNe energy input rate. (This circumstance can be important if 
subsequent observations will restrict from above such parameters of 
clumpuscules as ${\cal N}$, $M_c$, $v_c$.) It worth to stress
that since heating due to evaporating clumpuscules 
is much higher than radiation losses, it can support energy 
balance in halo on radial scales larger than $r_0$: clumpuscules with 
radial scale $r_0\simlt 10$ kpc can maintain radial distribution of halo 
gas with $R_h\sim 15$ kpc (though $r_0>10$ kpc seems more realistic). 
%

\bigskip
\centerline{\bf 8. Discussion and conclusions.}
\medskip
\par
The effects connected with mass loss by DM-clouds due 
to their collisions with each other and photoevaporation by ionizing 
radiation, are interesting not only by possible observational manifestation,
but also by their impact on dynamical and chemical evolution of the Galaxy. 
We discuss here only qualitative aspects of such influence -- detailed 
analysis requires a separate consideration. As shown in \S 7, the mass 
loss rate by a single cloud ionized by the background UV radiation is 
$\dot M_c\sim 10^{-13}~\msun$ yr$^{-1}$. Thus, a clumpuscule of 
$\sim 10^{-3}~\msun$ loses all its mass on the Hubble time.  
Moreover, the life time of DM-clouds can be 
even less than the Hubble time since they cross regions near hot stars with 
enhanced UV flux. In addition, it is almost certain that the background 
UV radiation can be more intense in young Galaxy due to violent star 
formation. If one assumes that all evaporated material is trapped by inner 
parts of the Galaxy ($R\simlt 15$ kpc), it follows inevitably that the total 
mass of luminous matter in the Galaxy must be comparable with (or higher 
than) that contained now in DM-clouds. Such a conclusion would obviously 
contradict observations, if mean surface density in the Galaxy due to 
DM-clouds is order of $\Sigma\simgt 10^2~\msun$ pc$^{-2}$ as estimated 
in [18]. One can escape this paradox by limiting from above the cloud 
radius $R_c$. Assuming clumpuscules to be gravitationally bound [6, 18], 
and with accounting Eq. (13) we obtain for cloud life time against 
photoevaporation $t_{ev} \simeq 1.2\times 10^{23}T_cR_c^{-1/2}$. Thus, 
clouds survive in the Hubble time $t_H$ if $R_c<1-3.5\times 10^{12}$ cm 
for cloud temperature $T_c=2.7$ and 5 K, respectively. (Note, that the 
restriction is weaker for higher temperature of DM-clouds: it scales 
as $\propto T_c^2$). One should stress, however, that the evaporating 
gas (as well as dispersed in cloud collisions) can in principle 
settle down in the outer Galaxy and form an extended ionized disk, as 
suggested first in [40]. 

A lower limit for cloud radius $R_c$ can be obtained if possible interaction  
of DM-clouds with the Solar system is taken into account. The frequency of 
DM-clouds crossing the Solar system is

$$\nu_S\sim \pi R_S^2 {\cal N}v_c,$$ 
where ${\cal N}\sim f/\pi R_c^2r_0$, $R_S\sim 40$ AU. This gives 

$$R_c\sim \sqrt{{fv_c\over \nu_S r_0}}R_S.$$
One can assume that $\nu_S\leq (100~{\rm yr})^{-1}$ --- more frequent collisions 
might result in detectable perturbations of planetary orbits. Then, for 
$r_0=10$ kpc and adopted parameters we get $R_c\geq 10^{11}$ cm. 

The net mass per unit area deposited by evaporating clumpuscules can be 
found as 

$$\Delta \Sigma=\Sigma[(1+2t_H/t_{ev})^2-1],$$ 
where $\Sigma$ is the present surface density in clumpuscules. Requiring 
$\Delta \Sigma<\Sigma_v$, where $\Sigma_v\sim 50~\msun$ pc$^{-2}$, we arrive 
at 

$$\Sigma<{\Sigma_v\over [(1+2t_H/t_{ev})^2-1]}.$$
For the lower limit $R_c=10^{11}$ cm this gives $\Sigma<\Sigma_v$. This 
estimate conflicts with the lower limit, $\Sigma> 100~\msun$ pc$^{-2}$,  
obtained in [18]. The contradiction can be avoided if the covering factor 
$f$ is two times smaller than estimated in [16] from the flux monitoring data.  
Note in this connection that in [18] $f\sim 10^{-4}$ is mentioned as a 
conservative value. [One should stress that since $2t_H/t_{ev}< 1$ and 
$t_{ev}\propto R_c^{-1/2}$ the obtained upper limit weakly depends on 
the parameters: approximately as $(f\nu_S r_0/v_c)^{1/4}$, though it is 
sensitive (inversely proportional) to the fraction of photoevaporated 
mass confined by the Galaxy.]

Transverse velocity of clouds should decrease proportionally 
$v_{c\perp}\propto R_c$ to keep the duration of extreme scattering events 
constant. This means in turn, that orbits of DM-clouds are strongly 
stretched, so that the radial component 
($\leq 500$ km s$^{-1}$) is much bigger than the transverse one. 

With such restrictions the current mass rate processed in cloud collisions, 
$\nu_GM_c\propto f^2R_c^{-1}$, can reach about $2-20~\msun$ yr$^{-1}$ and 
leads to overproduction of mass in the Galaxy. However, due to strong 
dependence on $f$ it can be as small as $<0.4-4~\msun$ yr$^{-1}$ for 
$f<10^{-3}$. 

Although at present the origin of DM-clouds is far from being understood 
well, one may say tentatively that they were born on prestellar stages of 
the universe, and are thus the most old objects [13]. This means that 
they might have primordial chemical composition, and being mixed with the 
interstellar gas decrease its metallicity. In other words, models of 
chemical evolution of the Galaxy which incorporate input of mass from 
photoevaporating and collisionally dispersed DM-clouds, should predict less 
efficient enrichment by metals than those without DM-clouds. It follows, 
in particular, that  since clouds concentrate to central 
regions of the Galaxy, it should make radial gradient of metallicity much 
softer than the observed one, or even to inverse it, (unless $r_0$ 
is large enough, $r_0>10 - 15$ kpc) -- otherwise, 
star formation rate must increase strongly in central regions to balance 
mass input with primeval composition. 

We summarise our findings: 

1) Collisions of dense gaseous clumpuscules responsible for the extreme 
scattering of quasar radio emission, are frequent enough: depending on 
their total number in the Galaxy around $10^3$ to $10^4$ collisions per year 
can take place. 

2) After collisions gas of clumpuscules is heated and then dissipates and 
mixes with the interstellar gas. The corresponding mass input rate can reach 
$1-10~\msun$ yr$^{-1}$. 
In this case their impact on dynamical and chemical evolution of the Galaxy 
can be important. In particular, it might result in softening of 
radial gradients of the stellar metallicity. 

3) On post-collisional stages shocked clumpuscules and their 
fragments appear as transient optical sources with temperature and
luminosity close to those of massive stars, the duration of these stages 
is about 10 days. 

4) On later stages expanding fragments can be seen in 21 cm, with duration 
of 1 Myr, and the total mass of such fragments in the Galaxy luminous in 
21 cm is of $3\times (10^5-10^6)~\msun$. This gas can be distinguished from 
the interstellar HI gas by its high velocity dispersion. 

5) DM-clouds surrounding 
hot stars form HII halos around them with electron density of 
$n_e\sim 0.65$ cm$^{-3}$, and can be seen in $H_\alpha$ with emission measure 
$EM\sim 20$ cm$^{-6}$ pc$^{-1}$ even for stars far from the galactic plane 
where interstellar gas is diluted. 

6) Ionized gas outflowing the atmospheres of DM-clouds 
can be a substantial mass source -- with the rate 
$2-20~\msun$ yr$^{-1}$ -- for the interstellar gas. If most of this 
evaporated gas is trapped by the Galaxy, clumpuscule 
radius can be constrained as $R_c\leq 3.5\times 10^{12}$ cm 
from requirement the mass deposited to be less than the total luminous mass.  
In turn, it leads to the conclusion that orbits of clouds are strongly 
stretched with small transverse velocity component. In this case, 
the surface mass density in DM-clouds can be as small as 
50 $\msun$ pc$^{-2}$. 

7) In general, gas lost by DM-clouds has high velocity relative to the ISM 
($\leq 500$ km s$^{-1}$ [18]). Subsequent mixing with the interstellar 
gas transfers its kinetic energy to heat and is an 
additional heating source of the ISM. This can explain large radial scale 
length of the halo gas ($\sim 15$ kpc) if the scale length 
of DM-clouds distribution (1) is high enough ($\simgt 10 $ kpc). 

Possible detection (or, contrary, non-detection at given threshold 
of sensitivity) of optical emission, 
including $H_\alpha$, and emission in 21 cm from clumpuscules will be 
of great importance for understanding the nature of dark matter and 
for more confident estimates (or firm constraints) of the contribution 
of dense molecular clumpuscules to the galactic mass. In principle, one
can expect that dense atomic clouds, suggested in [11] as possible dark
matter objects, also may manifest themselves in $H_\alpha$ and 21 cm
emissions through photoionization by UV background and collisions. However, 
their parameters (sizes, densities, velocity dispersion) are too 
uncertain now to make definite conclusions. In [7, 8, 12, 14, 41] baryonic dark
matter is assumed to form aggregates similar to globular clusters -- 
baryonic dark clusters -- with a fraction of baryons in gas clouds 
(see for detailed discussion [12, 14]). Stability of such clusters
implies that gas clouds are collisionless, however, even rare collisions
might provide substantial mass supply into the ISM if the clouds are numerous
enough. Also, photoevaporation due to external UV photons must be 
important for such clouds, however, effects from neighbouring (brown
dwarf) stars seem to be more influential [14]. 

I am grateful to P.M.W. Kalberla who brought my attention to 
the extreme scattering events problem and to the paper [18]. 
I thank D. Bomans, E. Corbelli, R.-J. Dettmar, P.M.W. Kalberla, 
R. Schlickeiser, A. Schr\"oer and 
C. Taylor for valuable discussion and comments. I also thank Ph. Jetzer
for comments on dark matter clusters. 
The work was supported by the National Programme ``Astronomy'' and by 
the Center ``Ground Based Astronomy'' under the National Programme 
``Integration'' (projects 352 and 353). I acknowledge the hospitality 
of the Osservatorio Astrofisico di Arcetri where this work has been
finished. 

\bigskip
\centerline{\bf REFERENCES }
\medskip
\ref {1.} Alcock C., Akerlof C.W., Allsman R.A., et al., Nature. 1993, 
{\bf 365}, 621 

\ref {2.} Auborg E., Bazeyre P., Brehin S., et al., Nature. 1993, 
{\bf 365}, 623 

\ref {3.} Alcock C., Allsman R. A., Alves D., et al., 
Astrophys. J. 1997, {\bf 486}, 697 

\ref {4.} Komberg B.V., Kompaneets D.A., Lukash V.N., Astron. Rept, 1995, 
{\bf 72}, 457 

\ref {5.} Pfenniger D., Combes F., Martinet L., Astron. \& Astrophys.
1994, {\bf 285}, 79

\ref {6.} Pfenniger D., Combes F., Ibidem. 1994, {\bf 285}, 94 

\ref {7.} De Paolis F., Ingrosso G., Jetzer Ph., Roncadelli M., 
Astron. \& Astrophys. 1995, {\bf 295}, 567

\ref {8.} De Paolis F., Ingrosso G., Jetzer Ph., Quadir A., Roncadelli M., 
Astron. \& Astrophys. 1995, {\bf 299}, 647

\ref {9.} Henriksen R.N., Widrow L.M., Astrophys. J. 1995, {\bf 441},
70

\ref {10.} De Paolis F., Ingrosso G., Jetzer Ph., Roncadelli M., in:
Dark Matter in Galaxies, eds. M. Persic and P. Salucci, ASP Conf. Ser.
117, 1997, 266

\ref {11.} Field G.B., Corbelli E., in:
Dark Matter in Galaxies, eds. M. Persic and P. Salucci, ASP Conf. Ser.
117, 1997, 258

\ref {12.} Gerhard O., Silk J., Astrophys. J. 1996, {\bf 472}, 34

\ref {13.} Pfenniger D., Combes F., 1998, astro-ph/9801319

\ref {14.} De Paolis F., Ingrosso G., Jetzer Ph., Roncadelli M., 
Astrophys. J. 1999, to appear, astro-ph/9801126

\ref {15.} Dixon D.D., Hartmann D.H., Kolaczyk E.D., Samimi J., Diehl
R., Kanbach G., Mayer-Hasselwander H., Strong A.W., New Astronomy. 
1998, astro-ph/9803237

\ref {16.} Fiedler R.L., Dennison B., Johnston K.J., Hewish A., 
Nature. 1987, {\bf 326}, 675 

\ref {17.} Romani R., Blandford R.D., Cordes J.M., Nature. 1987, {\bf 328}, 
324 

\ref {18.} Walker M., Wardle M., Astrophys. J., 1988, {\bf 498}, L125

\ref {19.} Draine B. T., Astrophys. J. 1998, submitted, astro-ph/9805083

\ref {20.} Binney J., Tremain S., {\it Galactic Dynamics},
Princeton Univ. Press: Princeton, 1987

\ref {21.} Stone M.E., Astrophys. J. 1970, {\bf 159}, 293 

\ref {22.} Hausmann M. A., Astrophys. J. 1981, {\bf 245}, 72 

\ref {23.} Gilden D. L., Astrophys. J. 1984, {\bf 279}, 335 

\ref {24.} Lattanzio J. C., Monagham J. J., Pongracic H., Schwarz M. P., 
Month. Not. RAS. 1985, {\bf 215}, 125 

\ref {25.} Klein R. I., McKee C. F., Woods D. T., in: The Physics 
of Interstellar and Intergalactic Medium, eds. A. Ferrara, C. Heiles,  
C. F. McKee, P. Shapiro, ASP Conf. Ser. 80, 1995, 366 

\ref {26.} Miniati F., Jones T. W., Ferrara A., Ryu D., Astrophys. J.
1997, {\bf 491}, 216 

\ref {27.} Zeldovich Ya.B., Raizer Yu.P., Physics of Shock waves and 
High Temperature Hydrodynamic Phenomena, London: Academic Press, 1965 

\ref {28.} Kalberla P.M.W., Kerp J., Astron. Astrophys. 
1998, {\bf 339}, 745

\ref {29.} Kaplan S.A., Pikelner S.B., Physics of the Interstellar Medium, 
Moscow, Nauka, 1979 

\ref {30.} Wolfire M.G., Hollenbach D., McKee C.F., Tielens A.G.G., 
Bakes E.L.O., Astrophys. J. 1995, {\bf 443}, 152 

\ref {31.} Klein R.I., McKee C.F., Collela P., Astrophys. J. 1994, 
{\bf 420}, 213 

\ref {32.} Shchekinov Yu.A., Astron. Astrophys. 1996, {\bf 314}, 927 

\ref {33.} Heiles C., Astrophys. J., 1997, {\bf 481}, 193

\ref {34.} Walker M., Wardle M., in: Proc. Stromlo Workshop on 
High Velocity Clouds. 1998, in press, astro-ph/9811209

\ref {35.} Dyson J.E., Astrophys. Space Sci. 1968, {\bf 1}, 388 

\ref {36.} Dove J.B., Shull J.M., Astrophys. J. 1994, {\bf 430}, 222 

\ref {37.} Kutyrev V.P., Reynolds R.J., Astrophys. J. 1989, 
{\bf 344}, L9 

\ref {38.} Spitzer L., Physical Processes in the Interstellar Medium,
     NewYork, Wiley, 1978 

\ref {39.} Hensler G., Samland M., Michaelis O., Severing I., in: 
The Physics of Galactic Halos, eds. H. Lesch, R.-J. Dettmar, U. Mebold, 
R. Schlickeiser, Akademie Verlag, Berlin. 1996, 225 

\ref {40.} Salpeter E. E., in: The Physics
of Interstellar and Intergalactic Medium, eds. A. Ferrara, C. Heiles,
C. F. McKee, P. Shapiro, ASP Conf. Ser. 80, 1995, 264

\ref {41.} Wasserman I., Salpeter E.E., Astrophys. J. 1994, {\bf 433},
670


\vfill\eject
\end